\documentclass[11pt]{article}
\usepackage{amsmath,amssymb, color,cite}

\usepackage{mathrsfs}
\usepackage{slashed}

\usepackage{amsfonts}
\usepackage{cleveref}

\makeatletter
\@addtoreset{equation}{section}
\makeatother

\newcommand{\be}{\begin{equation}}
\newcommand{\ee}{\end{equation}}
\newcommand{\bea}{\setlength\arraycolsep{2pt} \begin{eqnarray}}
\newcommand{\eea}{\end{eqnarray}}

\def\0{{\sst{(0)}}}
\def\1{{\sst{(1)}}}
\def\2{{\sst{(2)}}}
\def\3{{\sst{(3)}}}
\def\4{{\sst{(4)}}}
\def\5{{\sst{(5)}}}
\def\6{{\sst{(6)}}}
\def\7{{\sst{(7)}}}
\def\8{{\sst{(8)}}}
\def\sst#1{{\scriptscriptstyle #1}}

\allowdisplaybreaks

\begin{document}

\begin{flushright}
\hfill{}

\end{flushright}

\vspace{25pt}
\begin{center}
{\Large {\bf Higher derivative asymptotic charges and internal Lorentz symmetries}}

\vspace{35pt}
{\bf Mahdi Godazgar and George Macaulay }

\vspace{15pt}

{\it School of Mathematical Sciences,
Queen Mary University of London, \\
Mile End Road, E1 4NS, United Kingdom.}

 \vspace{35pt}

\date\today

\vspace{20pt}

\underline{ABSTRACT}
\end{center}

\noindent In line with a recent proposal for the study of asymptotic gravitational charges, we investigate higher derivative asymptotic charges. We show that the higher derivative BMS charges are related to the two-derivative BMS charges. Significantly, we find that internal Lorentz transformations are relevant in the higher derivative case in contrast to the two-derivative case. We give a prescription for their precise definition and derive the associated charges, finding, again, a relation with two-derivative BMS charges.

\noindent

\thispagestyle{empty}

\vfill
E-mails: m.godazgar@qmul.ac.uk, g.long@qmul.ac.uk

\pagebreak

\section{Introduction} \label{Intro}

Recently,  it has been argued that all gravitational charges, including asymptotic dual charges \cite{dual0, dualex}, may be investigated using the first order tetrad formalism, including any terms in the action that do not contribute to the equations of motion \cite{Godazgar:2020gqd}.  For example, asymptotic dual charges may be derived from the Holst term or the Nieh-Yan term in the presence of fermions.  

In this paper, we continue this investigation by considering the higher derivative terms that may be added to the action, namely the Pontryagin and Gauss-Bonnet terms, in the context of asymptotically flat spacetimes.  In Ref.\ \cite{Godazgar:2020gqd}, it was shown that asymptotic BMS charges derived from such terms would be subleading at null infinity.  Higher derivative terms have been considered in various settings in the literature \cite{Aros:1999id, Durka:2011yv, Durka:2011zf, Corichi:2013zza, Jacobson:2015uqa, Prabhu:2015vua, Frodden:2017qwh, Aneesh:2020fcr}.  While the importance of higher derivative terms has been stressed in much of the literature connected to the Anti-de Sitter case starting with Ref.\  \cite{Aros:1999id}, their relevance as far as defining asymptotic charges are concerned have largely\footnote{In Ref.\ \cite{Jacobson:2015uqa}, the contribution of the Gauss-Bonnet term at the bifurcate horizon of a black hole is important.} been dismissed, due to the fact that the charges are at subleading orders.  However, as argued in Refs.\ \cite{fakenews, dualex}, charges that appear at subleading orders may be important and ought not be dismissed.  The plethora of potentially observable gravitational memory effects \cite{Pasterski:2015tva, Mao:2018xcw, Mao:2020vgh, Seraj:2021qja, Seraj:2021rxd} and their relation to asymptotic symmetries and charges \cite{Strominger:2014pwa} makes it all the more important to investigate all the possible asymptotic charges. 

Another interesting aspect of higher derivative terms is the role of internal Lorentz symmetries.  In the first order tetrad formalism, the gauge symmetries of the theory is extended beyond diffeomorphisms to include the Lorentz transformations that one may apply to Lorentz indices.  This poses a dilemma, because we may have potentially enlarged our set of asymptotic symmetry, or improper gauge, generators; hence, changing the theory.  Some \cite{Jacobson:2015uqa, Prabhu:2015vua,Barnich:2016lyg, Barnich:2016rwk, DePaoli:2018erh, Barnich:2019vzx, Oliveri:2019gvm, Godazgar:2020gqd, Godazgar:2020kqd, Nguyen:2020hot} have argued that the asymptotic diffeomorphisms fix (or ought to fix) the asymptotic internal Lorentz transformations, rendering them trivial, while, others more recently \cite{Freidel:2020xyx, Freidel:2020svx, Freidel:2021fxf} have argued that internal Lorentz symmetries are physically non-trivial and do contribute to asymptotic charges.  In fact, they propose a further extension of the asymptotic symmetry algebra.  From a purely agnostic point of view,  the fact that putative internal Lorentz charges derived from the two-derivative (Palatini plus Holst) action are in any case trivial\footnote{This result relies on the Bondi-Sachs gauge.} \cite{Godazgar:2020kqd} makes this discussion rather moot.  Moreover, the treatment of the internal Lorentz symmetry does not affect the result for the diffeomorphism charges \cite{Oliveri:2020xls}.  Therefore, it is worth considering this issue for the higher derivative terms. 

Interestingly, we find that the asymptotic charges derived from asymptotic diffeomorphisms and internal Lorentz transformations are both related to expressions derived from the asymptotic diffeomorphisms for the two-derivative terms.  The result for the asymptotic diffeomorphisms relies on a judicious choice of splitting between integrable and non-integrable terms informed by their BMS transformation properties.  This provides further clues as to how one can define this splitting, which is in principle arbitrary.

Our results regarding the importance of internal Lorentz symmetries are in agreement with Ref.\ \cite{Freidel:2021fxf}.  Moreover, we give a prescription for handling internal Lorentz transformations in the asymptotic context.  Whereas in the case of global symmetries, the combined action of the diffeomorphisms and internal Lorentz transformations (the Kosmann derivative) on the vierbein is chosen so that it vanishes for Killing symmetries \cite{Jacobson:2015uqa}, for asymptotic symmetries, the action on the \emph{gauge-fixed} vierbein  must be chosen as to match the action of the diffeomorphisms on the \emph{gauge-fixed} metric.  The residual internal Lorentz transformations then correspond to the improper internal Lorentz transformations and are, therefore, physically relevant.

In section \ref{setup}, we review properties of asymptotically flat spacetimes and the necessary Einstein equations, as well as introducing the higher derivative terms that will be the focus of this paper. In section \ref{LorentzInvariance}, we derive the asymptotic symmetries of asymptotically flat spacetimes in the tetrad formalism.  This requires an analysis of the internal Lorentz symmetry and a judicious Lorentz gauge fixing of the zweibein.  We find that the asymptotic symmetry group is given by the BMS group, as well as another function on the 2-sphere that corresponds to residual internal Lorentz transformations.  In section \ref{BMSCharges}, we derive the BMS charges associated with the higher derivative terms, concentrating on BMS supertranslations.  In particular, we find that conserved quantities arise in the absence of Bondi news. In section \ref{LorentzCharges}, we derive the charges associated with the internal Lorentz symmetry generator.  In particular, these charges are integrable.  We conclude with some discussion in section \ref{sec:dis}.

\noindent \textbf{Notation:} Latin indices $(a, b,...)$ denote internal Lorentz indices and are raised and lowered with respect to the Lorentz metric $\eta_{ab}$. We use Greek letters $(\mu, \nu,...)$ to denote the spacetime indices.  The vierbein $e^a= {e^a}_{\mu} dx^{\mu}$. The spacetime metric can be expressed in terms of the vierbein as ${ds^2=g_{\mu \nu} dx^\mu dx^\nu= \eta_{ab} e^a e^b}$. Finally, we define the curvature 2-form as ${\mathcal{R}_{ab}(\omega)= d\omega_{ab} +\omega_{ac} \wedge {\omega^c}_b}$.  Indices $I,J,...$ will denote spacetime indices on the round 2-sphere and will be lowered and raised using the round 2-sphere metric $\gamma_{IJ}$ and its inverse, respectively, except where explicitly stated otherwise.

\section{Preliminaries} \label{setup}
\subsection{Gauge choices}
Let $(u,r,x^I=\{\theta,\phi\})$ be coordinates on our spacetime manifold such that the metric takes the form
\begin{equation} \label{metric}
ds^2 = -F e^{2\beta} du^2 -2e^{2\beta} du dr +r^2 h_{IJ} (dx^I-C^I du)(dx^J-C^J du)
\end{equation}
This is our diffeomorphism gauge choice. Here $r$ is a radial coordinate. The spacetimes we are considering here are asymptotically flat and we use the Bondi definition of asymptotic flatness where the metric functions obey the following fall-off conditions \cite{bondi,sachs}
\begin{align} \label{metricexpansions}
F(u,r,x^I) &= 1+\frac{F_0 (u,x^I)}{r}+\frac{F_1 (u,x^I)}{r^2}+o(r^{-2}) \notag\\
\beta(u,r,x^I) &= \frac{\beta_0(u,x^I)}{r^2}+o(r^{-2})  \notag\\
C^I(u,r,x^I) &= \frac{C_0^I(u,x^I)}{r^2}+\frac{C_1^I(u,x^I)}{r^3}+o(r^{-3}) \notag\\
h_{IJ}(u,r,x^I) &= \gamma_{IJ}+\frac{C_{IJ}(u,x^I)}{r}+\frac{C^2\gamma_{IJ}}{4r^2}+\frac{D_{IJ}(u,x^I)}{r^3}+o(r^{-3}),
\end{align}
where we have additionally assumed an analytic expansion to the order required for our calculations. Here $C^2 = C_{IJ} C^{IJ}$ and $\gamma_{IJ}$ is the round metric on the 2-sphere. We have assumed an expansion where the coefficient of $r^{-2}$ in the $h_{IJ}$ expansion has vanishing traceless part. Relaxing this assumption results in logarithmic terms appearing in the expansions after consideration of the Einstein equations, thus breaking our assumption of analyticity to this order. As a further consequence, the Weyl scalars do not satisfy the peeling property \cite{NP61}. There is residual gauge freedom allowing us to fix the radial distance by setting $\det h = \det \gamma$ which determines the trace of each term in the $h_{IJ}$ expansion. To this order, traces simply vanish, tr$C=$ tr$D=0$.

It is also a requirement for the calculations in this paper, that we explicitly pick a Lorentz gauge. Let our frame fields be
\begin{equation} \label{vierbein}
e^0 = \tfrac{1}{2} F du +dr, \quad e^1 = e^{2\beta} du \quad \text{and} \quad e^i = r E^i_I (dx^I - C^I du). \\
\end{equation}
where $E^i_I$ is a zweibein associated with the 2-metric $h_{IJ}$: $(h^{-1})^{IJ} E^i_I E^j_J= \eta^{ij}$ and $\eta_{ij} E^i_I E^j_J = h_{IJ}$. Contrary to all other fields, $I,J,...$ indices on $E^i_I$ will be raised and lowered with $h_{IJ}$. For more details, we refer the reader to Refs.\ \cite{fakenews, Godazgar:2020kqd}.

\subsection{Einstein Equations}  \label{EinsteinEquations}
The energy-momentum tensor for many physically relevant spacetimes has strong fall-off conditions and the expressions for the charges drastically simplify when assuming this to the appropriate order in the Einstein equations. In particular, for the components of interest for us here, one finds that \cite{fakenews}
\begin{align} \label{Einstein}
G_{00}=o(r^{-4}) \quad &\implies \quad \beta_0 = -\tfrac{1}{32} C^2\notag \\
G_{0i}=o(r^{-3}) \quad &\implies \quad C_0^I  = -\tfrac{1}{2} D_J C^{IJ}\notag \\
G_{ij}=o(r^{-4}) \quad &\implies \quad  \partial_u D_{IJ}  = \tfrac{1}{8}C_{IJ}\partial_u C^2 -\tfrac{1}{4}F_0 C_{IJ} -\tfrac{1}{2}D_{\langle I} C_{1 J \rangle} \notag \\
&\hspace{10mm}-\tfrac{1}{8} C_{IJ}D_K D_L C^{KL} +\tfrac{1}{32}D_{\langle I }D_{J\rangle} C^2 \notag \\
&\hspace{10mm}+\tfrac{1}{2} D_{\langle I} ( C_{J\rangle K} D_L C^{KL} ) -\tfrac{1}{8} D_{\langle I }C^{KL} D_{J\rangle} C_{KL}
\end{align}
where $D_K$ denotes a covariant derivative associated with the round 2-sphere metric $\gamma_{IJ}$ and angled brackets $\langle, \rangle$ on pairs of indices denote the symmetric, trace-free part.  We shall be assuming the above equations in this paper. 

\subsection{Higher derivative terms}
Refs.\ \cite{Godazgar:2020gqd,Godazgar:2020kqd} advocate the use of the first order tetrad formalism for defining gravitational asymptotic charges.  In the first order tetrad formalism the gravitational fields are given by the vierbein and spin connection: $\{ e^a, \omega^{ab}\}$, which are taken to be independent.  The Einstein tensor that contributes to the left-hand-side of the Einstein equation is then obtained by the inclusion of the Palatini term in the action
\begin{align} \label{Palatini}
I_P &= \frac{1}{32\pi} \int_{\mathcal{M}} \varepsilon_{abcd} \mathcal{R}^{ab}(\omega) \wedge e^c \wedge e^d.
\end{align}
More precisely, if this is the only contribution to the total action, one can show the equations of motion are the vacuum Einstein equations and the vanishing of torsion
\begin{equation}
T^a = d e^a + \omega^a{}_{b} \wedge e^b = 0.
\end{equation}
Applying the covariant phase space formalism \cite{peierls,berg,CWitten,Crknovic88,IW, WZ} to this action, one obtains \footnote{The slash on the variational symbol $\delta$ on the LHS is to indicate that the variation is not, in general, integrable.}  
\begin{align} \label{PalatiniCharge}
\slashed{\delta} \mathcal{Q}_{\xi} &= \frac{1}{16\pi} \varepsilon_{abcd} \int_{\partial \Sigma} \iota_{\xi} e^c \delta \omega^{ab} \wedge e^d,
\end{align}
which at leading order gives rise to the BMS charges and at lower orders, the subleading BMS charges \cite{fakenews}, when appropriate boundary conditions are assumed.

Adding another possible contribution to the action, the Holst term,\footnote{The parameter $\lambda_I$ is known as the Barbero-Immirzi \cite{BarberoG:1994eia, Immirzi:1996dr} parameter.}
\begin{align} \label{Nieh-Yan}
I_{H} &= \frac{i\lambda_I}{16\pi} \int_{\mathcal{M}}  \mathcal{R}_{ab}(\omega) \wedge e^a \wedge e^b
\end{align}
does not change the equations of motion.  However, the covariant phase space formalism does give  the dual BMS charges obtained in Refs.\ \cite{dual0, dualex}:
\begin{align} \label{Nieh-Yan Charge}
\slashed{\delta} \widetilde{\mathcal{Q}}_{\xi} &= \frac{1}{8\pi} \int_{\partial \Sigma} \iota_{\xi} e^a \delta \omega_{ab} \wedge e^b.
\end{align}

The inclusion of the Holst term in the action gives rise to important physics, namely the dual charges. It is therefore natural to consider other possible contributions to the action that do not affect the equations of motion and test whether these give rise to new charges in the same way as the Holst term does. The terms we consider here are the Gauss-Bonnet and Pontryagin terms given by
\begin{align} \label{PGBaction}
I_{GB} = \frac{1}{2}\varepsilon_{abcd} \int_{\mathcal{M}} \mathcal{R}^{ab} \wedge \mathcal{R}^{cd}  \quad \text{and} \quad   I_P &= \frac{i}{2} \int_{\mathcal{M}} \mathcal{R}^{ab} \wedge \mathcal{R}_{ab}
\end{align}
which can be combined as
\begin{align} \label{PGBaction2}
I &=P_{abcd} \int_{M} \mathcal{R}^{ab} \wedge \mathcal{R}^{cd} 
\end{align}
where $P_{abcd}=\frac{1}{2}\varepsilon_{abcd}+ \frac{1}{2}i \eta_{a[c}\eta_{d]b}$. The presymplectic potentials are given by
\begin{align} \label{PresymplecticFormPGB}
\theta_{GB} = \varepsilon_{abcd} \delta \omega^{ab} \wedge \mathcal{R}^{cd} \quad \text{and} \quad \theta_P=\delta\omega_{ab}\wedge \mathcal{R}^{ab}.
\end{align}

\section{Gauge transformations of the fields}  \label{LorentzInvariance}
In the metric formulation, we know that the metric transforms under diffeomorphisms $\xi$ that act on the metric via a Lie derivative
\begin{equation}
\delta_\xi g_{\mu \nu} = \mathcal{L}_\xi g_{\mu \nu}.
\end{equation}
Furthermore, we know that the set of diffeomorphisms $\{ \xi^{\mu} \}$ preserving the form of the metric \eqref{metric} form the BMS group \cite{bondi,sachs} and are given by 
\begin{equation}
\xi^u= f, \qquad \xi^r = \tfrac{r}{2} ( C^I \partial_I f -D_I \xi^I),  \qquad
\xi^I = Y^I -\int^{\infty}_r dr^{\prime} \frac{e^{2\beta}}{{r^{\prime}}^2}{(h^{-1})}^{IJ} \partial_J f 
\label{BMSgroup}
\end{equation}
with $f=s(x^I) + \tfrac{u}{2} D_I Y^I$. Here $s(x^I)$ are scalars on the 2-sphere parameterising  supertranslations and $Y^I(x^I)$ are conformal Killing vectors on the 2-sphere corresponding to the $SL(2,\mathbb{C})$ part of the BMS group.\footnote{In this paper, we shall restrict ourselves to supertranslations, i.e.\ set $Y^I =0$, as the supertranslations are the most novel feature of the BMS group.}  

Unlike the metric, the vierbein will transform under internal Lorentz transformations, as well as diffeomorphisms. In particular, 
\begin{equation}
\delta_{\xi,\Lambda} e_{\mu}^{a} = \mathcal{K}_{\xi,\Lambda} e_\mu^a= \mathcal{L}_\xi e_\mu^a + \Lambda^a{}_{b} e_\mu^b,
\end{equation}
where $\mathcal{K}_{\xi,\Lambda}$ denotes the Kosmann derivative, the combined effect of both types of gauge transformations.  For a general internal Lorentz transformation $\Lambda^{ab}$, this is consistent with the definition of the metric in terms of the vierbein. Thus we have an extra set of gauge parameters. 

However, when considering particular gauge transformations that correspond to Killing or asymptotic symmetries , the action of the diffeomorphism generators will act differently on the vierbein compared with the metric.  Thus, for consistency, we may wish to constrain the internal Lorentz generators.  Before considering the asymptotic case, we review the treatment in the fully covariant case.

\subsection{Kosmann derivative} \label{Kosmannderivative}
In the case where the theory remains fully covariant, i.e.\ where we have not fixed any of the diffeomorphism freedom, the only relevant generators are those that generate an isometry, defined by the condition that
\begin{equation}
\mathcal{L}_\xi g_{\mu \nu} = 0.
\end{equation}
Given such a Killing vector field, it is not only desirable, but necessary, in order to define Noether charges, that the Kosmann derivative acting on the vierbein along such a Killing direction vanishes too \cite{Jacobson:2015uqa}.  This is achieved by writing $\mathcal{K}_{\xi,\Lambda} e_\mu^a$ in terms of $\mathcal{L}_\xi g_{\mu \nu}$.  In particular, following \cite{Jacobson:2015uqa}, one can choose
\begin{equation}
\mathcal{K}_{\xi,\Lambda} e_\mu^a= \frac{1}{2} e^{a \nu} \mathcal{L}_\xi g_{\mu \nu}.
\end{equation}
This then fixes the Lorentz transformation generator $\Lambda$ in terms of the diffeomorphism generator $\xi$
\begin{equation}
\Lambda_{ab} = e^\mu{}_{[a} \mathcal{L}_\xi e_{b] \mu}.
\end{equation}
The fact that the Lorentz transformation generator is not independent implies that the Lorentz transformations introduced by the tetrad formalism are spurious; there ought not be any notions of symmetry or charges associated with their generators.  

Under Lorentz rotations $L^{ab}$ of the vierbein
\begin{equation}
e^a \rightarrow (Le)^a = L^a{}_b e^b,
\end{equation}
$\Lambda_{ab}$ transforms as a connection\footnote{Infinitesimally with $L = \delta + \lambda$, $$\Lambda^{(\lambda e)}
= - \mathcal{L}_\xi \lambda + [\lambda, \Lambda^{(e)}].$$}
\begin{equation} \label{Lambdaconn}
\Lambda^{(Le)}_{ab} \equiv (Le)^\mu{}_{[a} \mathcal{L}_\xi (Le)_{b] \mu} = L_{a}{}^c L_b{}^d \Lambda^{(e)}_{cd} + L_{[a}{}^c \mathcal{L}_{\xi} L_{b] c}
\end{equation}
so that the Kosmann derivative remains covariant under Lorentz transformations
\begin{equation} 
\mathcal{K}_{\xi, \Lambda^{(Le)}} (Le)_\mu^a = L^a{}_b \mathcal{K}_{\xi, \Lambda^{(e)}} e_\mu^b.
\end{equation}
Similarly, 
\begin{equation}
\mathcal{K}_{\xi, \Lambda} \omega_\mu^{ab} \longrightarrow  L^a{}_c  L^b{}_d \mathcal{K}_{\xi, \Lambda}  \omega_\mu^{cd}
\end{equation}
under a Lorentz transformation $L$.  This means that the Kosmann derivative can be used as a diffeomorphism and Lorentz covariant structure to construct charges. 

\subsection{Asymptotic symmetries in the tetrad formalism} \label{LorentzTransformations}
The discussion above assumes that one is interested in global symmetries.  In this case, the background diffeomorphism and Lorentz covariance may be retained and no further issues arise.  When dealing with asymptotic symmetries, however, the situation is more subtle.  In order to define asymptotic symmetries one must break the background diffeomorphism symmetry by fixing a particular coordinate system, such as Bondi coordinates, and likewise, the Lorentz symmetry by choosing a tetrad corresponding to the coordinate system. 

In Ref.\ \cite{Godazgar:2020gqd}, the criterion used to fix $\Lambda_{ab}$ was different to that elaborated above.  Having chosen the basis 1-forms \eqref{vierbein}, we require that the Kosmann derivative $\mathcal{K}_{\xi, \Lambda} e^a_\mu$ gives an expression that matches the variations of the metric components coming from $\mathcal{L}_\xi g_{\mu \nu}$.  In particular, we require an additional Lorentz transformation in order to restore the Lorentz gauge that we have fixed when choosing the null tetrad:
\begin{align}
& \mathcal{K}_{\xi, \Lambda} e^0_r =0  \quad \implies \quad \Lambda_{01}= - \partial_r \xi^r, \label{Lambda01} \\
& \mathcal{K}_{\xi, \Lambda} e^0_I = 0\quad \implies \quad \Lambda_{1 i} = \frac{1}{2r} E_{i}^{{\scriptscriptstyle I}} \left( F \partial_{{\scriptscriptstyle I}}\xi^u + 2 \partial_{{\scriptscriptstyle I}} \xi^r  \right),   \label{Lambda1i} \\
& \mathcal{K}_{\xi, \Lambda} e^1_I =  0 \quad \implies \quad \Lambda_{0i} = \frac{e^{2 \beta}}{r} E_{i}^{{\scriptscriptstyle I}} \partial_{{\scriptscriptstyle I}} \xi^u.   \label{Lambda0i} 
\end{align}
since we require that $\delta e^0_r = \delta e^0_I= \delta e^1_I=0$. Finally, the expression for $\Lambda_{ij}$ is derived by considering $\delta_{\xi,\Lambda} e^i_I =   r \delta_{\xi,\Lambda} E^i_I$ \footnote{To be clear, $\delta_\xi E^i_{I}$ represents an explicit transformation of the metric component functions that appear in $E^i_{I}$ using the prescription derived from $\delta_\xi g_{\mu \nu} =\mathcal{L}_\xi g_{\mu \nu}$.  In particular, $\delta_\xi \hat{E}^i_{I} = 0.$}
\begin{equation}  \label{Lambdaij} 
\Lambda_{ij} = \frac{1}{r} E^I{}_{[i} \mathcal{L}_\xi e_{j] I} - E^I{}_{[i} \delta_\xi E_{j] I},
\end{equation}
which can be expanded in orders of $1/r$, with \cite{Godazgar:2020gqd}
\begin{equation} \label{Lambdaijexp}
 \Lambda_{ij} = \gamma_{{\scriptscriptstyle I}{\scriptscriptstyle J}} \hat{E}^{{\scriptscriptstyle I}}_{[i} \mathcal{L}_{Y} \hat{E}^{{\scriptscriptstyle J}}_{j]} + O(1/r),
\end{equation}
where $\hat{E}^I_i$ are the zweibeins associated with the unit round-sphere metric $\gamma_{IJ}$ so that $\eta^{ij} \hat{E}_i^I \hat{E}_j^J = \gamma^{IJ}$. Clearly, the precise form of the subleading terms in \eqref{Lambdaijexp} will depend on how we choose to parametrise the zweibein $E^i_I$.  The transformation of the other components of the vierbein are consistent with the expressions for $\Lambda_{ab}$ given in equations \eqref{Lambda01}--\eqref{Lambdaij}.

To reiterate, the criterion for fixing $\Lambda_{ab}$ in the asymptotic case is very different to that used in the global case.  In the former case, the criterion is that once a vierbein has been chosen, we require that $\Lambda_{ab}$ be such as to keep it in its Lorentz frame.  However, we have kept a residual internal Lorentz symmetry in the $i$ direction.  Since we have (partially) fixed the Lorentz frame, an analysis of the transformation of $\Lambda_{ab}$ under an arbitrary Lorentz transformation $L_{ab}$ as we did in the global case is meaningless. However, the residual internal Lorentz symmetry in the $i$ direction means that we now need to consider Lorentz rotations of the zweibein:
\begin{equation} \label{resLor}
E^i_I \rightarrow L^i{}_{j} E^j_I.
\end{equation}
Under the transformation above, $\Lambda_{ab}$ for $ab \neq ij$ transforms as a tensor.  However, as is clear from equation \eqref{Lambdaij}, $\Lambda_{ij}$ will not transform as a connection ({\textit{cf.}} equation \eqref{Lambdaconn}).  Thus, the Kosmann derivative will not transform covariantly under Lorentz transformations acting on $i,j,\ldots$ indices.  In particular, this means that the BMS charge will depend on the choice of zweibein $E^i_I$.  

In summary, we find that breaking the diffeomorphism invariance in order to define the background of interest means that Lorentz covariance cannot be maintained and must too be broken.  The higher derivative charges will then depend on the choice of internal Lorentz gauge.  Our choice  for the vierbein is given in \eqref{vierbein}.  Furthermore, we require that
\begin{equation} \label{zweibeingauge}
E^i_I = X_I{}^J \hat{E}^i_J.
\end{equation}
where $X_{IJ}(u,r,x^I)$ is a tensor on the 2-sphere. While this choice may seem arbitrary, one advantage is that all expressions remain tensors on the 2-sphere, i.e.\ one does not break diffeomorphisms along the 2-sphere directions.  

Thus, we have two sets of asymptotic symmetry generators: the BMS generators \eqref{BMSgroup} and a residual internal Lorentz transformation that corresponds to internal Lorentz transformations of the round 2-sphere zweibein $\hat{E}^i$, which we shall denote by
\begin{equation} \label{lamij}
\lambda_{ij}= \lambda (x^I)\,  \varepsilon_{ij}.
\end{equation}
Note that the gauge choice \eqref{zweibeingauge} means that $\lambda$ depends only on the coordinates on the 2-sphere.  A different gauge choice would lead to an internal Lorentz transformation parameter $\lambda (u,r,x^I)$.

\section{Gauss-Bonnet and Pontryagin BMS Charges} \label{BMSCharges}
In this section, we consider the set of charges arising from the Gauss-Bonnet and Pontryagin terms generated by the BMS generators \eqref{BMSgroup}. In general, the charge variations resulting from the higher derivative terms are of the form \cite{Godazgar:2020gqd}
\begin{align}  \label{PandGBcharge}
\slashed{\delta}\mathcal{Q}_P = \int_S \delta \omega^{ab} \wedge \mathcal{K}_{\xi, \Lambda} \omega_{ab} \quad \text{and} \quad \slashed{\delta}\mathcal{Q}_{GB} = \varepsilon_{abcd} \int_S  \delta \omega^{ab} \wedge  \mathcal{K}_{\xi, \Lambda} \omega^{cd}.
\end{align}
We will now derive these expressions in the asymptotically flat case with the asymptotic symmetry generators corresponding to supertranslations. 

 \subsection{Derivation of $\Lambda_{ij}$} \label{Variationofthemetric}
First, we need to derive $\Lambda_{ij}$, defined by equation \eqref{Lambdaij}, to lower orders than that given in equation \eqref{Lambdaijexp}.  This expression will depend on the gauge choice \eqref{zweibeingauge}.  This gauge choice combined with the condition  $E^I_i E^J_j \eta^{ij} = (h^{-1})^{IJ}$ implies that 
\begin{eqnarray} \label{Xeqn}
{X^I}_K {X^J}_L \gamma^{KL} = (h^{-1})^{IJ}.
\end{eqnarray}
Viewed as matrices, this is just a change of basis.  Solving this equation for $X$ as a $r^{-1}$ expansion given that
\begin{equation}
(h^{-1})^{IJ} = \gamma^{IJ}-\tfrac{1}{r}C^{IJ} +\tfrac{1}{4r^2}C^2 \gamma^{IJ}-\tfrac{1}{r^3}D^{IJ}  + o(r^{-3})
\end{equation}
gives
\begin{eqnarray}
X_{IJ} &=& \gamma_{IJ}-\tfrac{1}{2r}C_{IJ} +\tfrac{1}{16r^2}C^2 \gamma_{IJ}+\tfrac{1}{r^3}\big{(}\tfrac{1}{32}C^2 C_{IJ} -\tfrac{1}{2}D_{IJ}\big{)}+o(r^{-3})
\end{eqnarray}

Now, we can derive $\Lambda_{ij}$ to the order required for the calculation.  Using \eqref{zweibeingauge}, equation \eqref{Lambdaij} reduces to
\begin{equation}
\Lambda_{ij} = \gamma_{IJ} \hat{E}_{[i}^{I}\, \xi^K \partial_K \hat{E}_{j]}^{J} + \tfrac{1}{2} \varepsilon_{ij} \epsilon^I{}_{J} (\partial_I\xi^J- C^J \partial_I \xi^u) + \tfrac{1}{2} \varepsilon_{ij} \epsilon^{JK} X^I{}_{J} (\xi^\mu \partial_\mu X_{IK} - \delta_\xi X_{IK}),
\end{equation}
where $\varepsilon_{23}=i$ and $\epsilon_{\theta\phi}=\sin \theta$.\footnote{The fact that $\varepsilon_{23}$ is imaginary is a consequence of the fact that we have chosen a complex null tetrad on the 2-sphere.}  Note that $\delta_\xi X_{IK}$ denotes the variation of the metric components in $X_{IK}$ derived from the equation $\delta_{\xi} g_{\mu \nu} = \mathcal{L}_{\xi} g_{\mu \nu}$.  Thus, \cite{fakenews}
\begin{align} \label{metvariations}
\delta_{\xi} F_0 &= s\partial_u F_0 -\tfrac{1}{2} \partial_u C^{IJ} D_I D_J s -D_I \partial_u C^{IJ} D_Js \notag\\
\delta_{\xi}  C_1^I &= s \partial_uC_1^I  +\tfrac{1}{16}\partial_u C^2 D^I s +F_0 D^I s -\tfrac{1}{4} C^{JK} D^I D_J D_K s -\tfrac{1}{2} C^{IJ} D_J \Box s \notag \\
&\hspace{10mm}+ \tfrac{1}{2} D^J C^{IK} D_J D_K s-\tfrac{3}{4} D^I C^{JK} D_J D_K s -\tfrac{1}{2}D_J C^{JK} D_K D^I s \notag\\
&\hspace{10mm}-\tfrac{1}{2}D^I D^J C_{JK} D^K s+\tfrac{1}{2} D^J D_K C^{KI} D_J s-C^{IJ} D_J s\notag \\
\delta_{\xi} C_{IJ} &=s\partial_u C_{IJ} -2D_{\langle I} D_{J \rangle} s\notag\\
\delta_{\xi}  C^2 &= s\partial_u C^2 -4C^{IJ} D_I D_J s\notag \\
\delta_{\xi} D_{IJ} &=s\partial_u D_{IJ}-2C_{1\langle I} D_{J \rangle} s -\tfrac{1}{4} C_{IJ} C^{KL} D_{K} D_{L} s-\tfrac{1}{8} C^2 D_{\langle I} D_{J \rangle} s \notag \\
&\hspace{10mm}+\tfrac{1}{8} D_{\langle I } C^2 D_{J\rangle} s +D_K C^{KL} C_{L\langle I} D_{J \rangle} s
\end{align}
Denoting the expansion \footnote{The previous analyticity assumptions made up to this point mean $\Lambda_{23}$  automatically has an analytic expansion to this order.}
\begin{eqnarray}
\Lambda_{23}= \Lambda_{23}^{(0)}+\frac{\Lambda_{23}^{(1)}}{r}+\frac{\Lambda_{23}^{(2)}}{r^2}+o(r^{-2})
\end{eqnarray}
we find
\begin{eqnarray}
 \Lambda_{23}^{(0)} &=& \hat{E}_{I [2} \mathcal{L}_Y \hat{E}^I_{3]} \nonumber \\
 \Lambda_{23}^{(1)} &=& V^I D_I f  \nonumber \\
 \Lambda_{23}^{(2)} &=& \tfrac{1}{2} D^I f (i D^J \widetilde{C}_{IJ}-V^J C_{IJ})
 \end{eqnarray}
where $V^I = \tfrac{1}{2} (\hat{E}_3^J D^I \hat{E}_{2J}-\hat{E}_2^J D^I \hat{E}_{3J})$.  The leading order term matches with equation \eqref{Lambdaijexp}, as expected.

\subsection{Charge Expansion} \label{ChargeExpansion}
This puts us in a position to obtain an expression for the Pontryagin and Gauss-Bonnet charge expansions in the form
\begin{align}
\slashed{\delta}\mathcal{Q_P} = \int_S d\Omega \hspace{2mm}\bigg{\{} \frac{{ (\slashed{\delta}\mathcal{I_P})_0 }}{r^2} +\frac{{ (\slashed{\delta}\mathcal{I_P})_1 }}{r^3} +\frac{{ (\slashed{\delta}\mathcal{I_P})_2 }}{r^4}+o(r^{-4}) \bigg{\}}
\end{align}
and
\begin{align}
\slashed{\delta}\mathcal{Q_{GB}} = \int_S d\Omega \hspace{2mm}\bigg{\{} \frac{{ (\slashed{\delta}\mathcal{I_{GB}})_0 }}{r^2} +\frac{{ (\slashed{\delta}\mathcal{I_{GB}})_1 }}{r^3} +\frac{{ (\slashed{\delta}\mathcal{I_{GB}})_2 }}{r^4}+o(r^{-4}) \bigg{\}}.
\end{align}
Note that it can be deduced almost immediately from \eqref{PandGBcharge} by considering the fall-off of each term in the sum and noting each $\mathcal{K}_{\xi, \Lambda} \omega^{ab}_I$ has a fall-off at least as fast as $\delta \omega^{ab}_I$ for a general variation, that the leading order terms are $\mathcal{O}(r^{-2})$ in the charge expansions. In the following calculations, we will assume the energy-momentum tensor obeys the fall-off conditions $T_{00}=o(r^{-5})$, $T_{0i}=o(r^{-3})$ and $T_{ij}=o(r^{-4})$ so we can use the Einstein equations \eqref{Einstein}. A calculation then shows
\begin{align}
 (\slashed{\delta}\mathcal{I_{GB}})_0 = 0  \hspace{10mm} \text{and} \hspace{10mm}  (\slashed{\delta}\mathcal{I_{P}})_0=0.
 \end{align}
This result holds in the full BMS case, i.e.\ for $Y^I \neq 0$.
 
\subsection{Gauss-Bonnet at $\mathcal{O}(r^{-3})$} \label{Gauss-BonnetO3}
At the next order, we obtain the following expression for the Gauss-Bonnet charge
\begin{align}
(\slashed{\delta}\mathcal{I_{GB}})_1 &= \delta \big{(}3D_{[I} s D^K C_{J]K} D^I D_L C^{JL}\big{)}  \notag\\
& + s\big{(} 2\big{(}  \partial_u D_{IJ} \partial_u \delta C^{IJ}-\delta D_{IJ} \partial_u^2 C^{IJ}\big{)} +\tfrac{1}{4} \big{(} \partial_u C^2 \delta F_0-\delta C^2 \partial_u F_0 \big{)} \notag \\
&+3\big{(}\partial_u C_1^I D^J \delta C_{IJ}-\delta C_1^I D^J \partial_u C_{IJ}  \big{)}+\tfrac{1}{8} \big{(}\partial_u C^2 D_I D_J \delta C^{IJ}-\delta C^2 D_I D_J \partial_u C^{IJ}\big{)} \notag\\
&+\tfrac{1}{16}\big{(}D_I \partial_u C^{IJ} D_J \delta C^2-D_I \delta C^{IJ} D_J \partial_u C^2 \big{)} \notag \\
&+ \tfrac{3}{2} D_I C^{IJ} \big{(}  \delta C_{JK} D_L \partial_u C^{KL}-\partial_u C_{JK} D_L \delta C^{KL}\big{)} \notag\\ 
&+ \tfrac{1}{32}\big{(}\delta C^2 \partial_u^2 C^2 -\partial_u C^2 \partial_u \delta C^2\big{)}+\tfrac{1}{8} \big{(}\delta(C^2 C_{IJ}) \partial_u^2 C^{IJ} - \partial_u(C^2 C_{IJ})\big{)}  \partial_u\delta C^{IJ}  \big{)} \notag  \\
&+D_I s \big{(}3F_0 D_J \delta C^{IJ}+ \tfrac{3}{2}D_J C^{JK} (D_K D_L \delta C^{IL}- D^I D^L \delta C_{KL})-3\delta C_{1J} \partial_u C^{IJ} \notag \\
&-4 C_{1J} \partial_u \delta C^{IJ} + \tfrac{5}{4} D_J C^2 \partial_u \delta C^{IJ}+\tfrac{5}{16} D_J \delta C^2 \partial_u C^{IJ} -\tfrac{1}{2} \partial_u C^{KL}  \delta(C^{IJ}D_K  C_{JL}) \notag \\
& -\tfrac{1}{8} \partial_u C^2 D_J \delta C^{IJ} +\tfrac{1}{2}C^{IJ}D_K \delta C^{KL}\partial_u C_{JL}- 2C^{IJ} D_K C_{JL}\partial_u \delta C^{KL} \notag \\
&+ \delta C_{JK}D_L C^{KL}\partial_u C^{IJ}\big{)} \notag\\
&-D_I D_J s \big{(} \delta F_0 C^{IJ}+\tfrac{1}{2} D_K( C^{IJ} D_L \delta C^{KL} ) +\tfrac{1}{8}(\delta C^2 \partial_uC^{IJ}-C^{IJ}\partial_u\delta C^2) \big{)} \notag\\
&-\tfrac{1}{2} D_K D_I D_J s C^{IJ} D_L \delta C^{KL} \notag\\
&+D_I (\Box +2) s \big{(}3\delta {C_1}^I -\tfrac{3}{2} \delta C^{IJ} D^K C_{JK} -\tfrac{3}{2} C^{IJ} D^K \delta C_{JK} -\tfrac{1}{16} D^I \delta C^2 \big{)} \notag \\
&+\tfrac{1}{8} \Box (\Box+2)s \delta C^2 .
\end{align}
It should be noted the choice of separation into the integrable and non-integrable piece is somewhat arbitrary. While, Wald and Zoupas \cite{WZ} give a formulation for a canonical way of doing this at the leading order, there is no such general formalism at subleading orders. The choice of separation here is because it leads to non-trivial charges. The integrable piece is conserved when the non-integrable piece vanishes for $\delta = \delta_{\xi,\Lambda}$. Focusing on the non-integrable piece and using the expressions for the metric variations \eqref{metvariations}, as well as Einsteins equations \eqref{Einstein}, we find
\begin{align}
({\slashed{\delta}\mathcal{I_{GB}})_1}^{(non-int)} &=3D_{ [ I}s D_{J]} \Box s D^I D_K C^{JK} \notag \\
&+ 3D_{[I} \big{(} D^K (s\partial_u C_{J] K} ) D^J D_L C^{IL}+D^K C_{J]K} D^J D_L (s\partial_u C^{IL})\big{)}
\end{align}
For a general $C_{IJ}$, the first and second line cannot cancel each other, so we need to consider them individually. The first line is zero when $s(x^I)=Y_{\ell,m}(x^I)$, a spherical harmonic, since then $\Box s = -\ell(\ell+1) s$ and the term vanishes since $D_{ [ I}s D_{J]}  s=0$. The second line vanishes only when $\partial_u C_{IJ}=0$. Hence, we have a set of charges
\begin{align}
{\big{(}\mathcal{Q_{GB}}\big{)}_1^{\ell,m}}^{(int)} = 3\int_S d \Omega \bigg{(}D_I Y_{\ell,m} D^K C_{JK} D^{[I} D_L C^{J]L}\bigg{)}
\end{align}
which are conserved in the absence of Bondi news, $\partial_u C_{IJ}=0$.

\subsection{Pontryagin at $\mathcal{O}(r^{-3})$} \label{PontryaginO3}
A similar calculation can be performed for the Pontryagin charge
\begin{align}
(\slashed{\delta}\mathcal{I_P})_1 &=\delta \big{(} -\tfrac{3}{2}D_I s F_0 D_J \widetilde{C}^{IJ}\big{)} \notag \\
&+ s\big{(} \big{(}\partial_u \widetilde{D}_{IJ}\partial_u \delta C^{IJ} - \delta \widetilde{D}_{IJ} \partial_u^2 C^{IJ} \big{)} +\tfrac{3}{2} \big{(}\delta C_1^I D^J \partial_u \widetilde{C}_{IJ}-\partial_u C_1^I D^J \delta \widetilde{C}_{IJ} \big{)} \notag\\
&+\tfrac{3}{32}\big{(} D_I \delta \widetilde{C}^{IJ} D_J \partial_u C^2 - D_I \partial_u \widetilde{C}^{IJ} D_J \delta C^2\big{)}\notag\\
&+\tfrac{1}{2}D_I C^{IJ} \big{(}\delta \widetilde{C}_{JK}D_L \partial_u C^{KL}-\partial_u \widetilde{C}_{JK}D_L \delta C^{KL}\big{)}\notag\\ 
&+\tfrac{1}{8}\big{(} D_I \partial_u C^{IJ} \delta(\widetilde{C}^{KL}D_J C_{KL})-D_I \delta C^{IJ} \partial_u(\widetilde{C}^{KL}D_J C_{KL})\big{)}-\tfrac{1}{32} \delta \widetilde{C}^{IJ} \partial_u C_{IJ} \partial_u C^2 \notag\\ 
&+\tfrac{1}{16}C^2 \big{(} \delta \widetilde{C}^{IJ} \partial_u^2 C_{IJ}- \partial_u \widetilde{C}^{IJ} \partial_u \delta C_{IJ}\big{)}+\tfrac{3}{32}\widetilde{C}^{IJ} \big{(}\delta C^2  \partial_u^2 C_{IJ} -\partial_u \delta C_{IJ} \partial_u C^2\big{)} \notag\\
&+ D_I s \big{(} \tfrac{3}{2}\delta F_0 D_J \widetilde{C}^{IJ}+2C_{1J} \partial_u \delta \widetilde C^{IJ} +\tfrac{3}{2} \delta C_{1J} \partial_u \widetilde{C}^{IJ} +\tfrac{3}{4} D_J \delta C^{IJ} D_K D_L \widetilde{C}^{KL} \notag \\
 &-\tfrac{1}{2}\delta C_{JK} D_L C^{KL} \partial_u \widetilde{C}^{IJ}+\tfrac{1}{8}\partial_u C^{IJ} \delta \big{(} \widetilde{C}_{KL}D_J C^{KL}\big{)} -\tfrac{1}{8}D_J \delta C^{IJ} \widetilde{C}_{KL}\partial_u C^{KL}\notag\\
 &+\tfrac{1}{2}\partial_u \delta C^{IJ} \widetilde{C}_{KL} D_J C^{KL}-\tfrac{3}{8}D_J C^2 \partial_u \delta \widetilde{C}^{IJ} -\tfrac{3}{32} D_J \delta C^2 \partial_u \widetilde{C}^{IJ} \big{)} \notag\\ 
 &+D_I D_J s \big{(} \tfrac{1}{4} D_K C^{IJ} D_L \delta \widetilde{C}^{KL} +\tfrac{1}{8}C^{IJ}\delta \big{(} \widetilde{C}^{KL} \partial_u C_{KL} \big{)} +\tfrac{1}{16} \delta C^2 \partial_u \widetilde{C}^{IJ}\big{)}\notag\\
& +\tfrac{1}{4}D_K D_I D_J s C^{IJ} D_L \delta \widetilde{C}^{KL}\notag\\
&+D_I (\Box+2)s \big{(} \tfrac{3}{2} \epsilon^{IJ} C_{1J} -\tfrac{7}{32} \epsilon^{IJ}D_J \delta C^2 +\tfrac{3}{8}\delta\big{(} C^{JK} D^I \widetilde{C}_{JK} \big{)} \big{)}
\end{align}
Once again, there is freedom in the choice of separation, but with this choice, the integrable piece corresponds to a conserved charge under sensible conditions on the metric and non-trivial $s(x^I)$. Using \eqref{metvariations} and \eqref{Einstein} , we obtain
\begin{align}
({\slashed{\delta}\mathcal{I_{P}})_1}^{(non-int)} &=\tfrac{3}{2}F_0 \epsilon_{IJ} D^I \Box s D^J s\notag\\
&+\tfrac{3}{2}F_0 D^I s D^J \big{(}s\partial_u \widetilde{C}_{IJ}\big{)} +\tfrac{3}{8} s D_I \widetilde{C}^{IJ} D_J s\partial_u C^{KL} \partial_u C_{KL}\notag\\
& -\tfrac{3}{4} D_I \widetilde{C}^{IJ} D_J s D_K D_L \big{(} s \partial_u C^{KL} \big{)}
\end{align}
Again, the first line is zero when $s(x^I)=Y_{\ell,m}(x^I)$, a spherical harmonic. For the second and third line to vanish, we require the absence of Bondi news, $\partial_u C_{IJ}=0$. Hence, we obtain a set of charges
\begin{align}
{\big{(}\mathcal{Q_{P}}\big{)}_1^{\ell,m}}^{(int)} =-\frac{3}{2}\int_S d \Omega \bigg{(}D_I Y_{\ell,m}  F_0 D_J  \widetilde{C}^{IJ} \bigg{)}
\end{align}
which are again conserved when $\partial_u C_{IJ}=0$.

\subsection{Physics of the charges} \label{PhysicsofCharges}
If one writes the Kerr metric in Bondi coordinates \cite{godazgar2019dual} , it can be shown that $C_{IJ}=0$ and hence the charges are each trivially zero. Doing the same with the Taub-NUT metric
\begin{equation}
ds^2=-f(r)(dt+2\ell \cos \theta d \phi)^2 +f(r)^{-1} dr^2 +(r^2+\ell^2) (d\theta^2 +\sin^2 \theta d\phi^2)
\end{equation}
where $f(r)=\frac{r^2-2mr+\ell^2}{r^2+\ell^2}$, we obtain non-vanishing expressions for the charges when $s$ is an axisymmetric spherical harmonic ($m=0$) and $\ell=2L>0$, given by
\begin{align}
{\big{(}\mathcal{Q_{GB}}^1\big{)}_{2L,0}}^{(int)} &= 48\sqrt{\pi (4L+1)} \ell^2 \notag\\
{\big{(}\mathcal{Q_{P}}^1\big{)}_{2L,0}}^{(int)} &=24\sqrt{\pi (4L+1)} m\ell
\end{align}
where $\ell$ and $m$ in these expressions are the Taub-NUT parameters. It is unsurprising that the charges vanish for Kerr and are non-trivial for Taub-NUT as we expected these charges to encompass information about the topology of the spacetime. Evidently, they contain information about the NUT-parameter.

\subsection{Supertranslation invariance of the charges} \label{SupertranslationInvariance}
The charges derived above with the conditions $\partial_u C_{IJ}=0$ (absence of Bondi news) and choice of $s=Y_{\ell,m}$ are conserved, but we have only demonstrated supertranslation invariance of $\mathcal{Q_{GB}}^{\ell,m}$ and $\mathcal{Q_{P}}^{\ell,m}$ for the supertranslation generated by $s=Y_{\ell,m}$. If one considers the variation of the charges for general $s$, even with the vanishing of the Bondi news tensor, the expression obtained is non-zero. However, it is possible to modify the charges, by adding an extra term each, so that the new expressions enjoy full supertranslation invariance. Let
\begin{align} \label{altGBcharge}
{\big{(}\mathcal{Q_{GB}}\big{)}_1}^{(int)} &\rightarrow 3\int_S d \Omega \bigg{(}D_I s D^K C_{JK} D^{[I} D_L C^{J]L}+sD^K C_{JK} D_I D^{[I} D_L C^{J]L}\bigg{)} \\
{\big{(}\mathcal{Q_{P}}\big{)}_1}^{(int)} &\rightarrow -\frac{3}{2}\int_S d \Omega \bigg{(}D_I sF_0 D_J  \widetilde{C}^{IJ} +s D_I F_0 D_J \widetilde{C}^{IJ}\bigg{)} \label{altPcharge}
\end{align}
The additional terms are each zero for the Taub-NUT spacetime so the previous discussion is unaffected. The addition of the new terms can be justified by adding their negative counterparts to the non-integrable pieces. Now for any spherical harmonic $s(x^I)$, in the absence of Bondi news, applying \eqref{metvariations}, we find
\begin{align}
{\big{(}\slashed{\delta} \mathcal{Q_{GB}}\big{)}_1}^{(non-int)} &= 3\int_S d \Omega\hspace{1mm} D_J (\Box+3) s D_I \big{(}s D^{[I} D_K C^{J]K}\big{)}  \\
{\big{(}\slashed{\delta}\mathcal{Q_{P}}\big{)}_1}^{(non-int)} &= \frac{3}{2}\int_S d \Omega\hspace{1mm} \epsilon_{IJ} D^I (\Box+3) s D^J \big{(}sF_0\big{)}
\end{align}
where the Ricci identity has been used. In both cases, the non-integrable piece can be written as the sum of a total derivative and a second term which is zero by the torsion free property of $D_I$.
\begin{align} \label{altGBnon-int}
{\big{(}\slashed{\delta} \mathcal{Q_{GB}}\big{)}_1}^{(non-int)} &= 3\int_S d \Omega\hspace{1mm} D_I \bigg{(} D_J (\Box+3) s  \big{(}s D^{[I} D_K C^{J]K}\big{)} \bigg{)}\\ \notag
&\hspace{30mm} -D_{[I} D_{J]} (\Box+3)s  \big{(}s D^{I} D_K C^{JK}\big{)}   \\
{\big{(}\slashed{\delta}\mathcal{Q_{P}}\big{)}_1}^{(non-int)} &= \frac{3}{2}\int_S d \Omega\hspace{1mm} D^J \bigg{(} \epsilon_{IJ} D^I (\Box+3) s \big{(}sF_0\big{)}\bigg{)}- \epsilon_{IJ} D^J D^I (\Box+3) s  \big{(}sF_0\big{)} \label{altPnon-int}
\end{align}
If the total derivative term can be ignored, then we have shown the charges are conserved in the absence of Bondi news. It can be shown that even for the Taub-NUT spacetime, the total derivative terms are zero for any regular $s$. The alternative charge expressions \eqref{altGBcharge} and \eqref{altPcharge} are therefore conserved in the absence of Bondi news for any supertranslation parameter $s(x^I)$, provided this condition is met. Furthermore, it can be shown these expressions are invariant under the transformation generated by any supertranslation parameter. A demonstration of this amounts to replacing one $s(x^I)$ in \eqref{altGBnon-int} and \eqref{altPnon-int} with another regular arbitrary function on the 2-sphere and observing that both expressions still vanish.

It is possible to argue these expressions for the charges are a more sensible choice given this property. It is also interesting to observe from \eqref{altGBcharge} and \eqref{altPcharge} that if total derivatives can be ignored in the integrand, then the charges can be written as
\begin{align}
{\big{(}\mathcal{Q_{GB}}\big{)}_1}^{(int)} &= -\frac{3}{2}\int_S d  \Omega \hspace{1mm} s \big{(}D_I D_J \widetilde{C}^{IJ}\big{)}^2 \\
{\big{(}\mathcal{Q_{P}}\big{)}_1}^{(int)} &= \frac{3}{2}\int_S d \Omega  \hspace{1mm} s F_0 \big{(}D_I D_J \widetilde{C}^{IJ}\big{)}
\end{align}
We observe that the integrand in the Gauss-Bonnet charge is proportional to the square of the integrand of the two-derivative dual BMS charge and the integrand of the Pontryagin charge is proportional to the product of the two-derivative BMS and dual BMS charges.

\section{Gauss-Bonnet and Pontryagin Lorentz Charges} \label{LorentzCharges}
In section \ref{BMSCharges}, we derived the BMS charges associated with the Gauss-Bonnet and Pontryagin charges.  As we found in section \ref{LorentzTransformations}, in addition to the BMS symmetry generators, there is also an internal Lorentz symmetry generator parameterised by a function $\lambda(x^I)$ (see equation \eqref{lamij}), which corresponds to the freedom in defining the zweibein corresponding to the round 2-sphere metric.  

Under the action of this residual internal Lorentz transformation, the spin connection transforms as 
\begin{gather}
\delta_{\lambda} \omega_{01} = 0, \quad
\delta_{\lambda} \omega_{0i} = \lambda {\varepsilon_i}^j \omega_{0j} , \quad
\delta_{\lambda} \omega_{1i} = \lambda {\varepsilon_i}^j \omega_{1j}  \notag \\
\delta_{\lambda} \omega_{ij} =-d\lambda \varepsilon_{ij} +\lambda( {\varepsilon_i}^k \omega_{kj} -{\varepsilon_j}^k \omega_{ki} ).
 \label{connectionLorentz}
\end{gather}
We evaluate the charge variations corresponding to the Gauss-Bonnet and Pontryagin contributions in the action. Using the expressions for the presymplectic potential in \eqref{PresymplecticFormPGB} and the expressions above, we are able to obtain charge variations associated with this Lorentz gauge symmetry. The charges obtained are in fact integrable with
\begin{align}
\mathcal{Q_P}^{Lorentz} &=  \int_S   \hspace{1mm}  \lambda \varepsilon^{ij} \mathcal{R}_{ij}\\
\mathcal{Q_{GB}}^{Lorentz} &=  4\int_S   \hspace{1mm}  \lambda \mathcal{R}_{01}.
\end{align}
Then, using the asymptotic expansions in \eqref{metricexpansions} and the Einstein equations \eqref{Einstein}, we find
\begin{align}
\mathcal{Q_P}^{Lorentz} &=  \int_S  d\Omega \hspace{1mm}  \lambda \bigg{\{} \big{(}4F_0 - \tfrac{1}{2} \partial_u C^2\big{)}r^{-1} + o(r^{-1}) \bigg{\}} \label{Lorentz:p}\\
\mathcal{Q_{GB}}^{Lorentz} &= \int_S  d\Omega \hspace{1mm} \lambda \bigg{\{} \big{(}2D_I D_J \widetilde{C}^{IJ} + C^{IJ}\partial_u \widetilde{C}_{IJ} \big{)}r^{-1} + o(r^{-1}) \bigg{\}} \label{Lorentz:gb}
\end{align}
Note the striking resemblance between these expressions and the leading order charges coming from the two-derivative action \cite{Godazgar:2020kqd}.  In fact, up to flux terms there is a precise match between these sets of charges.\footnote{Compare equation \eqref{Lorentz:p} with equation (3.5) of \cite{fakenews} and equation \eqref{Lorentz:gb} with equation (5.5) of \cite{dualex}.}  This suggests that the internal Lorentz charges are physically meaningful.

\section{Discussion} \label{sec:dis}

In this paper, we have focused on the higher-derivative Gauss-Bonnet and Pontryagin contributions to the action in the first order formalism, which do not affect the Einstein equations. We considered two sets of asymptotic symmetries: the BMS symmetry and the residual internal Lorentz symmetry. 

The charges corresponding to the BMS symmetry group arising from the Gauss-Bonnet and Pontryagin contributions to the action are combinations of the already known BMS and dual charges \cite{godazgar2019dual}. It is difficult to comment on what would happen at higher orders of $r^{-1}$ in these charge expansions, but it is likely that any further charges would also be combinations of charges coming from the two-derivative terms in the action. Perhaps the most prominent point of discussion here is the choice of separation into the integrable and non-integrable piece. The choice here is such that the integrable piece can be made to vanish for physically reasonable conditions on the metric with the resulting charges invariant under all supertranslations. This result suggests that requiring the non-integrable terms to vanish in the absence of flux, as well as requiring the resulting integrable charge to be BMS invariant is a reasonable criterion for distinguishing between the integrable and non-integrable pieces of the charge variation.  More generally, the relation between non-integrability and conservation requires more investigation.  There are indications that these two concepts need not be related in general, even though they are for BMS charges \cite{Ciambelli:2021nmv}.

The residual internal Lorentz symmetry generators lead to non-trivial charges, in contrast to the two-derivative case.  This suggests that in general, the internal Lorentz symmetry that appears in the tetrad formulation of GR is important and should not be ignored.  Furthermore, interestingly, as with the BMS charges, the internal Lorentz charges correspond to the charges coming from the two-derivative terms in the action.

The implications of these results for the first law of black hole mechanics is an interesting subject that we hope to address in the future.

\section*{Acknowledgements}
We would like to thank Hadi Godazgar and Malcolm Perry for useful discussions.
M.G.\ is supported by a Royal Society University Research Fellowship and G.M.\ is supported by a Royal Society Enhancement Award.

\bibliographystyle{utphys}
\bibliography{NP}

\end{document}